# On the deprojection of the Galactic bulge

James Binney[1] and Ortwin Gerhard[2]
[1] *Theoretical Physics, Keble Road, Oxford OX1 3NP*
[1] *Astronomisches Institut, Universität Basel, Venusstrasse 7, CH-4102 Binningen, Switzerland*

**ABSTRACT**
An algorithm is developed and tested for the problem posed by photometric observations of the bulge of the Milky Way. The latter subtends a non-trivial solid angle on the sky, and we show that this permits inversion of the projected brightness distribution under the assumption that the bulge has three orthogonal mirror planes of specified orientation. A serious error in the assumed orientation of the mirror planes should be detectable.

## 1 INTRODUCTION

In the last few years evidence has been accumulating that the galactic bulge is barred (e.g. Blitz 1993; Gerhard 1995). The distribution of near-infrared light from the direction of the galactic centre provides the most direct evidence for non-axisymmetry of the bulge since at $\lambda \sim 2\,\mu$m starlight is the principal contributor to the overall surface brightness at ($|l| \lesssim 20°, |b| \lesssim 8°$), and at such wavelengths our view of this region is only moderately obscured by dust.

Blitz & Spergel (1991) first detected tell-tale signs of the bulge's triaxiality in the $2\,\mu$m data of Matsumoto et al (1982), but these old data were far too noisy to permit detailed model-building. The near-infrared brightness distribution determined by the DIRBE experiment aboard the *COBE* satellite (Weiland et al 1994) are of very much higher quality and several groups are currently basing models of the bulge on these data. This activity raises the following question: how uniquely does the projected brightness distribution of the bulge determine the three-dimensional luminosity density in the bulge? We address this question. We are primarily interested in the feasibility, in principle, of the inversion of ideal data, rather than the practical question of whether any given body of data is of sufficient quality to be usefully inverted.

Since the bulge's projected brightness distribution is a function of the two coordinates of the sky, while the luminosity density of a bar is a function of all three spatial coordinates, it would be natural to assume that the latter cannot be unambiguously determined from the former. We show, however, that it is possible in many cases to recover accurately the three-dimensional luminosity distribution from the projected density provided the bulge is known to have three mutually orthogonal mirror planes of known orientation. Thus the ambiguity inherent in determination of the luminosity density of the bulge can be largely reduced to that involved in choosing the orientation of the bulge's mirror planes. Moreover, it turns out that the algorithm we develop can offer valuable guidance as to the orientation of these planes.

In Section 2 we formulate a Richardson–Lucy (Richardson 1972; Lucy 1974) deprojection algorithm for the recovery of bulge models from surface-brightness data, and explain how the assumption that the bulge is eight-fold symmetric about given axes enables a three-dimensional model to be inferred from two-dimensional data. In Section 3 we test the performance of this algorithm on pseudo-data. Section 4 sums up.

## 2 A RICHARDSON–LUCY ALGORITHM

### 2.1 Motivation

As Contopoulos (1954) already showed, the elliptical image of a distant galaxy can be interpreted as the projection of an infinite number of different elliptical bodies (see also Stark 1977). In particular, the image of any body whose isodensity surfaces are self-similar ellipsoids can be interpreted as the image of an axisymmetric body. It is helpful to understand why the non-negligible angular size of the Milky Way's bulge may allow less freedom in the interpretation of its photometry, and in particular betray the bulge's triaxiality.

From photometry alone, one cannot tell from what distance one receives any given contribution to the brightness observed at some point on the sky. Consequently, one can deform a galaxy into a different object with the same image by moving stars along their lines of sight. (In general the luminosities of the stars will have to be adjusted as they move nearer to or further from the observer.) When a galaxy is viewed from afar, all lines of sight are parallel and such reorganizations of the galaxy can easily carry it between two states of symmetry, such as ellipsoidal brightness distributions.

When a galaxy subtends a non-trivial solid angle, it is not clear that the system can be deformed between states of symmetry without changing the projected image. In particular, if we wish a deformation to be compatible with the system retaining mirror symmetry in three orthogonal planes, motion of any star along its line of sight will have to be accompanied by the motion of its mirror-image stars in directions that generally do not run parallel to their lines of sight. Thus such deformations of the three-dimensional figure of the galaxy would be accompanied by deformation of the image. The mathematical scheme below exploits this connection to infer a mirror-symmetric distribution from a projected image.

### 2.2 Mathematical formulation

We employ coordinate axes which coincide with the principal axes of the bulge. Then associated with any point $x \neq \mathbf{0}$, there are eight distinct points $x_i$, $i = 1, \ldots, 8$, symmetrically placed with respect to the principal axes, such that $\nu(x_i) = \nu(x)$, where $\nu(x)$ is the luminosity density due to stars at $x$. We refer to the $x_i$ as the "symmetric companions" of $x$.

The assumed octant-symmetry of the bulge may be expressed mathematically by

$$\nu(x) = \int_{+\text{octant}} \mathrm{d}^3 x' \, \delta_8(x, x') \nu(x'), \tag{1}$$

where

$$\delta_8(x, x') \equiv \delta(|x_1| - x'_1)\delta(|x_2| - x'_2)\delta(|x_3| - x'_3). \tag{2}$$

Notice that $\delta_8$ vanishes unless its second argument lies in the positive octant, $x_i \geq 0$.

The surface brightness in the direction from the Sun $\boldsymbol{\Omega}$ is

$$I(\boldsymbol{\Omega}) = \int_0^\infty \mathrm{d} s \, \nu\bigl(x(s)\bigr), \tag{3}$$

where $x(s) \equiv x_\odot + s\boldsymbol{\Omega}$ is the point distance $s$ from the Sun in the direction $\boldsymbol{\Omega}$. Substituting from (1) into (3), we have

$$\begin{aligned} I(\boldsymbol{\Omega}) &= \int \mathrm{d}^3 x' \nu(x') \int_0^\infty \mathrm{d} s \, \delta_8\bigl(x(s), x'\bigr) \\ &= \int_{+\text{octant}} \mathrm{d}^3 x' \nu(x') K(\boldsymbol{\Omega}|x'), \end{aligned} \tag{4a}$$

where

$$K(\boldsymbol{\Omega}|x') \equiv \int_0^\infty \mathrm{d} s \, \delta_8\bigl(x(s), x'\bigr). \tag{4b}$$

The kernel $K$ does not satisfy the normalization condition $\int \mathrm{d}^2 \boldsymbol{\Omega}\, K = 1$ required of a probability distribution. In fact

$$\begin{aligned} \int \mathrm{d}^2 \boldsymbol{\Omega}\, K(\boldsymbol{\Omega}|x') &= \int \mathrm{d}^2 \boldsymbol{\Omega} \int \mathrm{d} s \, \delta_8\bigl(x(s), x'\bigr) \\ &= \int \mathrm{d}^3 x \, \frac{1}{s^2(x)} \delta_8(x, x') \\ &= \begin{cases} 0 & \text{for } x' \text{ not in +octant,} \\ \displaystyle\sum_{i=1}^8 \frac{1}{s^2(x'_i)} \equiv \frac{1}{\bar{s}^2(x')} & \text{otherwise.} \end{cases} \end{aligned}$$

Here $x'_i$ is the position of the $i^{\text{th}}$ symmetric companion of $x'$ and $s(x') \equiv |x' - x_\odot|$ is the distance from the Sun to $x'$. In view of (4) we define

$$\widetilde{K}(\boldsymbol{\Omega}|x') \equiv \bar{s}^2(x') K(\boldsymbol{\Omega}|x') \quad \text{and} \quad \widetilde{\nu}(x') \equiv \frac{\nu(x')}{\bar{s}^2(x')}. \tag{5}$$

$\widetilde{K}(\boldsymbol{\Omega}|x)$ obviously vanishes unless the direction $\boldsymbol{\Omega}$ points to one of the $x_i$. When $\boldsymbol{\Omega}$ does so point, $\widetilde{K}$ is $\delta$-function like.

Integrating $\widetilde{K}$ over small a solid angle $\Delta\boldsymbol{\Omega}_i$ around $\boldsymbol{\Omega}_i$, we have

$$\begin{aligned} \int_{\Delta_i} \mathrm{d}^2 \boldsymbol{\Omega} \, \widetilde{K}(\boldsymbol{\Omega}|x') &= \bar{s}^2(x') \int_{\Delta_i} \mathrm{d}^2 \boldsymbol{\Omega} \, K(\boldsymbol{\Omega}|x') \\ &= \bar{s}^2(x') \int_{\Delta_i} \mathrm{d}^2 \boldsymbol{\Omega} \int \mathrm{d} s \delta_8\bigl(x(s), x'\bigr) \\ &= \bar{s}^2(x') \iint_{\Delta_i} \frac{\mathrm{d}^2 \boldsymbol{\Omega} s^2 \mathrm{d} s}{s^2(x_i)} \delta_8\bigl(x(s), x'\bigr) \\ &= \frac{\bar{s}^2(x')}{s^2(x'_i)} \\ &= \frac{1}{\left(\sum_{j=1}^8 1/s^2(x'_j)\right) s^2(x'_i)} \simeq \tfrac{1}{8}. \end{aligned}$$

It follows that

$$\widetilde{K}(\boldsymbol{\Omega}|x) = \begin{cases} \bar{s}^2(x) \displaystyle\sum_{i=1}^8 \frac{\delta(\boldsymbol{\Omega} - \boldsymbol{\Omega}_i)}{s^2(x_i)} & \text{for } x \text{ in +octant} \\ = 0 & \text{otherwise.} \end{cases} \tag{6}$$

In conclusion, our Richardson–Lucy scheme for the inversion of equation (3) is

$$\begin{aligned} I_r(\boldsymbol{\Omega}) &= \int \mathrm{d}^3 x \, \widetilde{\nu}_r(x) \widetilde{K}(\boldsymbol{\Omega}|x) \\ &= \int \mathrm{d}^3 x \, \nu_r(x) K(\boldsymbol{\Omega}|x) \\ &= \int_0^\infty \mathrm{d} s \, \nu_r\bigl(x(s)\bigr), \end{aligned} \tag{7}$$

as expected, and

$$\begin{aligned} \widetilde{\nu}_{r+1}(x) &= \widetilde{\nu}_r(x) \int \mathrm{d}^2 \boldsymbol{\Omega} \, \widetilde{K}(\boldsymbol{\Omega}|x) \frac{I(\boldsymbol{\Omega})}{I_r(\boldsymbol{\Omega})} \\ &= \widetilde{\nu}_r(x) \int \mathrm{d}^2 \boldsymbol{\Omega} \, \bar{s}^2(x) \sum_{i=1}^8 \frac{\delta(\boldsymbol{\Omega} - \boldsymbol{\Omega}_i)}{s^2(x_i)} \frac{I(\boldsymbol{\Omega})}{I_r(\boldsymbol{\Omega})}. \end{aligned} \tag{8}$$

Or, simply

$$\nu_{r+1}(x) = \nu_r(x) \bar{s}^2(x) \sum_{i=1}^8 \frac{I(\boldsymbol{\Omega}_i)}{I_r(\boldsymbol{\Omega}_i)} \frac{1}{s^2(x_i)}. \tag{9}$$

The convergence properties of any R–L scheme may be improved by modifying the scheme so that the 'data' are not $I(\boldsymbol{\Omega})$ but $g(\boldsymbol{\Omega})I(\boldsymbol{\Omega})$, where $g(\boldsymbol{\Omega})$ is an arbitrary positive function (Lucy 1994; Dehnen 1995). The iterates $I_r$ produced by the modified scheme are such that $gI \ln gI_r$ rather than $I \ln I_r$ is maximized as $r \to \infty$. Multiplying both sides of equation (4a) by $g$, it is clear that the modified scheme has kernel $gK$, and it is easy on to show that for this kernel (9) becomes

$$\nu_{r+1}(x) = \nu_r(x) \sum_{i=1}^8 \frac{I(\boldsymbol{\Omega}_i)}{I_r(\boldsymbol{\Omega}_i)} \frac{g(\boldsymbol{\Omega}_i)}{s^2(x_i)} \bigg/ \sum_{i=1}^8 \frac{g(\boldsymbol{\Omega}_i)}{s^2(x_i)}. \tag{10}$$

In the experiments described below, the minimization of the residuals on the sky was usefully accelerated by taking $g$ to be of the form $g = I^{-\beta}$ with $\beta \geq 0$; five iterations with $\beta = 0.2$ were followed by four iterations with $\beta = 0.8$. The early iterations emphasize the residuals near the centre, while the later ones concentrate on diminishing the residuals further out.

## 3 TESTS OF THE ALGORITHM

We have tested the effectiveness of the deprojection algorithm given by (10) as follows. For each point of an $80 \times 80$ grid of points in the $(l, b)$ plane we calculate the line-of-sight integral $I \equiv \int ds\, \nu$ through a simple ellipsoidal density distribution $\nu(\mathbf{x})$. For most of our experiments $\nu$ was of the form

$$\nu(\mathbf{x}) = (x_e y_e y_e)^{-1} \exp\left[-\left(\frac{r_c^2 + x^2}{x_e^2} + \frac{y^2}{y_e^2} + \frac{z^2}{z_e^2}\right)^{1/2}\right], \quad (11)$$

where $r_c = 175$ pc is comparable with the spatial resolution of our calculation (see below). The line of sight from the Sun to $\mathbf{x} = 0$ is inclined at angle $\theta_0$ to the $(x, y)$ plane, and the $x$-axis is inclined at angle $\phi_0$ to the projection of the Sun-centre line onto the $(x, y)$-principal plane. We adopt $|\mathbf{x}_\odot| = 8$ kpc and integrate from $s = 2 - 14$ kpc down each line of sight. Subsequently, the value of $I$ at an arbitrary point $(l, b)$ is inferred by bilinear interpolation between the values taken by $\ln I$ on this sky-grid.

Our model density distribution is defined by the value of $\nu_r$ at each point on a Cartesian grid of $25 \times 25 \times 20$ points that covers the positive octant uniformly out to $x_{\max} = 4.2$ kpc, $y_{\max} = 3.2$ kpc, $z_{\max} = 2$ kpc. We refer to the region $|x| \leq x_{\max}$, $|y| \leq y_{\max}$, $|z| \leq z_{\max}$ as the model's "bounding box." The model may have a different orientation from that of the true distribution $\nu$; the Sun-centre line is inclined at angle $\theta$ to the model's equatorial plane, and the projection of the Sun-centre line into that plane is inclined at angle $\phi$ with respect to the model's $x$-axis.

For each point $(l, b)$ of the sky-grid we project $\nu$ by integrating along the section of the line of sight that lies within the bounding box. When performing $\int ds\, \nu_r$, values of $\nu_r$ are obtained by tri-linear interpolation of $\ln \nu_r$ on the three-dimensional grid.

The sky-grid is chosen to be sufficiently large that it covers all points at which the projected density is non-zero. Since the bounding box projects to an irregular shape on the sky — its silhouette; see Fig. 1 — the sky-grid inevitably contains many lines of sight which do not intersect the bounding box. Along such lines of sight the projected density is set to an arbitrary small value.

At points on the sky which lie near the boundary of the box's silhouette, the projected density is artificially small because only a small portion of the line of sight intersects the bounding box. Allowance must be made for this when comparing with either real data or our pseudo-data, which are obtained by integrating over the most significant 12 kpc of every line of sight. We make this allowance by projecting our initial model, which is given by an analytic function such as (11), twice, once along 12 kpc of each line of sight and once through the bounding box only. The ratio $I_0^t$ and $I_0^b$ between these two estimates is recorded and subsequently used to estimate $I_r$ through

$$I_r = \frac{I_0^t}{I_0^b} I_r^b, \quad (12)$$

where $I_r^b$ is the projection of $\nu_r$ through the bounding box. The best way of seeing that our R–L algorithm will reduce the residuals in $I/I_r$ with $I_r$ obtained in this way is to ob-

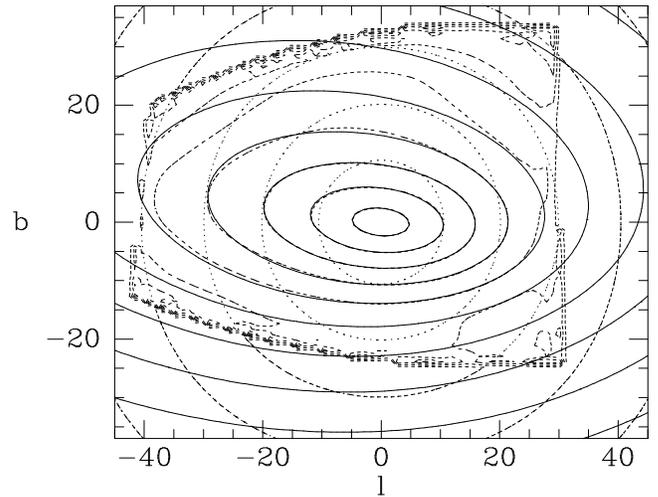

**Fig. 1** A trial at general orientation: $\theta_0 = \theta = 7°$, $\phi_0 = \phi = 18°$. After nine iterations an initially spherical model (dotted curves) has been deformed into a reasonable approximation (dashed curves) to the true surface-density contours (full curves). The contours of $I_9$ crowd at the edge of the silhouette of the bounding box because outside this silhouette $I_9$ is identical with $I_0$.

serve that one can write

$$\frac{I}{I_r} = \frac{\left(I_0^b/I_0^t\right)I}{I_r^b}. \quad (13)$$

That is, the factor in brackets on the top of the right-hand side of (13) is used to estimate the fraction of the data $I$ that arises in the bounding box, and then the R–L algorithm can be considered to be applied to projection from within the bounding box only.

Fig. 1 shows the result of an experiment in which the orientations of the trial and true distributions were general but identical: $\theta_0 = \theta = 7°$, $\phi_0 = \phi = 18°$. [These values are motivated by the work of Burton & Liszt (1978), who find $\theta \gtrsim 7°$, and of Binney et al. (1991), who find $\phi \simeq 18°$.] The true distribution had principal axes $(x_e, y_e, z_e) = (0.5, 0.3, 0.15)$ kpc, while the first trial distribution was spherical. The full and dotted curves show, respectively, the true and first-guess isophotes. The dashed curves show the isophotes of $I_9$. Near the centre these isophotes closely follow the true isophotes, but further out material deviations from the true isophotes are apparent, especially above the origin. The silhouette of the bounding box is delineated by the sharp crowding of the isophotes of $I_9$ where the transition is made from the region in which they have been adjusted by the R–L algorithm to that in which they coincide with the original (dotted) isophotes.

This sharp interface between the regions of modified and unmodified surface brightness arises because the initial guess (of a spherical distribution) was poor, with the result that the ratio $I_0^t/I_0^b$ used in equation (12) is sub-optimal. In such a case it would be immediately apparent from the data that the true distribution is somewhat flattened, and one would adopt a flattened distribution for $\nu_0$. Fig. 2 is the same as Fig. 1 but with $\nu_0$ given by equation (11) with $x_e = y_e = 0.4$ kpc, $z_e = 0.15$ kpc. The silhouette of the bounding box is now much less clear and the significant deviations between dashed and full isophotes are now confined to the upper left corner of the image, where the effects of truncation by the bounding box are most severe.

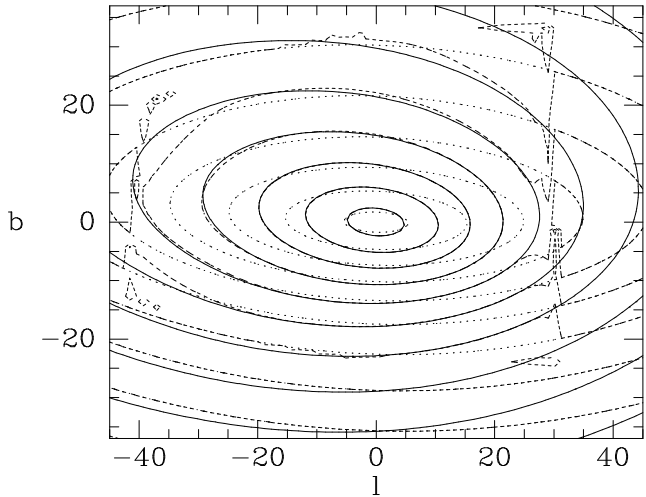

**Fig. 2** The same as Fig. 1 except that the initial distribution was flattened rather than spherical [$\nu_0$ was given by (11) with $x_e = y_e = 0.4\,\mathrm{kpc}$, $z_e = 0.15\,\mathrm{kpc}$]. In consequence the correction factors $I_0^t/I_0^b$ are more accurate and the silhouette of the bounding box is less apparent.

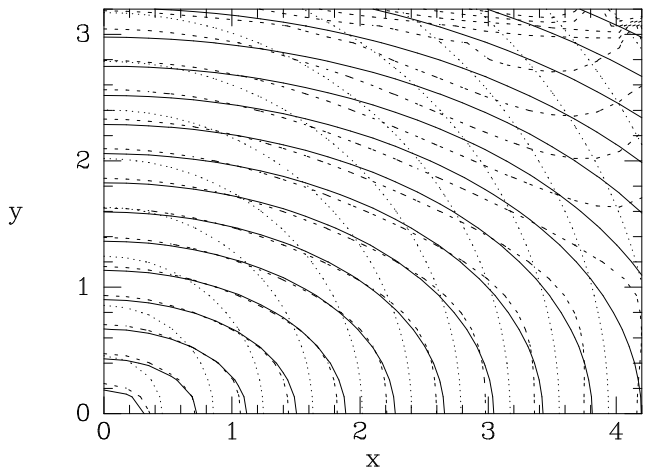

**Fig. 3** The density distributions that underlie the trial of Fig. 2. The full curves show the intersections of the true isodensity surfaces with the $(x, y)$ plane, while the dotted and the dashed curves show the same curves for $\nu_0$ and $\nu_9$.

Fig. 3 shows, for the inversion of Fig. 2, the contours in which isodensity surfaces of the true model (full contours), the initial-guess model $\nu_0$ (dotted contours) and the iterated model $\nu_9$ cut the $(x, y)$ plane. The innermost half of the dashed contours follow the full contours quite accurately. Thus the algorithm has correctly deduced that the bulge is bar-shaped. Further out, the dashed contours follow the full contours for $x \lesssim 3\,\mathrm{kpc}$ but then turn upwards towards the boundary. This effect arises from the need indicated above, to increase the model's surface brightness in the upper left of Fig. 2. In particular, if the iterations are started from an over-elongated distribution, such as (11) with axes $x_e = 0.6\,\mathrm{kpc}$, $y_e = 0.3\,\mathrm{kpc}$, $z_e = 0.15\,\mathrm{kpc}$, truncation by the box has the opposite effect, and the dashed curves in a plot such as Fig. 3 lie below the full curves at large $x$. Fig. 4 illustrates this point by showing the same physical region as is shown in Fig. 3 taken from a computation which employed a larger

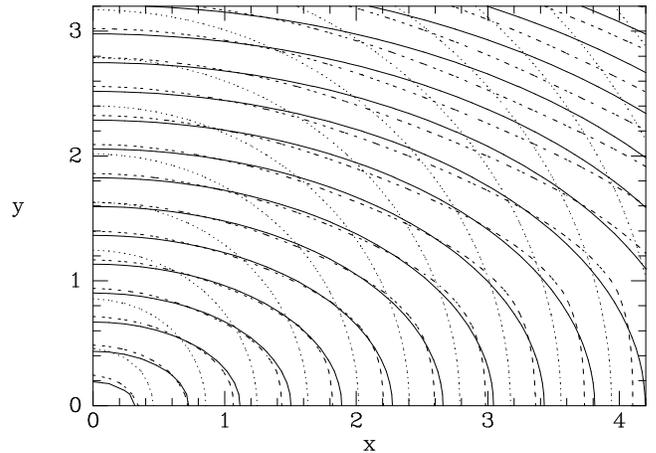

**Fig. 4** The same as Fig. 3 except that the bounding box has been extended to 6 kpc, 4.6 kpc and 2.9 kpc in $x$, $y$ and $z$, respectively, with a corresponding increase in the number of grid points within the box. The same physical region is shown.

bounding box. The spurious effects of the bounding box have now moved outwards.

Fig. 5 shows the result of a trial in which the principal axes of the density distribution that generated the data were rotated in the $(x, y)$ plane with respect to the assumed symmetry axes of the deprojected model. Specifically, $\theta_0 = \theta = 7°$, $\phi_0 = 30°$, $\phi = 18°$. Fig. 6 shows that the algorithm now has difficulty diminishing the residuals beyond a certain point. Moreover, in Fig. 5 the residuals form a clear pattern; the predicted and true isophotes cross one another three times in each revolution of the projected centre. Clearly, one could search for the true orientation of the bar by finding the orientation at which the residuals on the sky are smallest and have no clear angular pattern.

From Figs 1–5 we conclude that for $\theta_0 = \theta = 7°$, our R–L algorithm is able to infer the true brightness distribution of a triaxial bulge to good accuracy. Spurious effects generated by the bounding box are readily diagnosed and can be used to guide one's initial guess towards one that lies sufficiently close to the truth to give good final results. Alternatively, they can be shifted out of the region of astrophysical interest by using a larger bounding box. When the assumed symmetry axes are inclined with respect to the true ones, the residuals on the sky develop a pattern which suggests that the model needs to be re-oriented. Similar results are obtained for other values of $\theta$ and $\phi$, provided neither angle is chosen such that the Sun-centre line lies almost in a principal plane.

Fig. 7, which shows the result of a trial with $\theta_0 = \theta = 0$ and $\phi_0 = \phi = 18°$, shows that the scheme performs less well when the Sun-centre line lies close to the equatorial plane. Fig. 6 shows that the residuals in the $(l, b)$ plane decline at exactly the same rate as in the test shown in Fig. 2, yet along a line in Fig. 7 that is inclined at $\sim 18°$ to the $x$-axis, the dashed curves, which mark the intersection of the plane with the equidensity surfaces of $\nu_9$, turn abruptly downwards. As the inclination $\phi$ of the model to the Sun-centre line is varied, the inclination of the line on which the dashed curves have shoulders follows $\phi$. The sharpness of the shoulders gradually reduces as $\theta$ is increased. It seems that the algorithm correctly infers that $\nu$ is elongated along

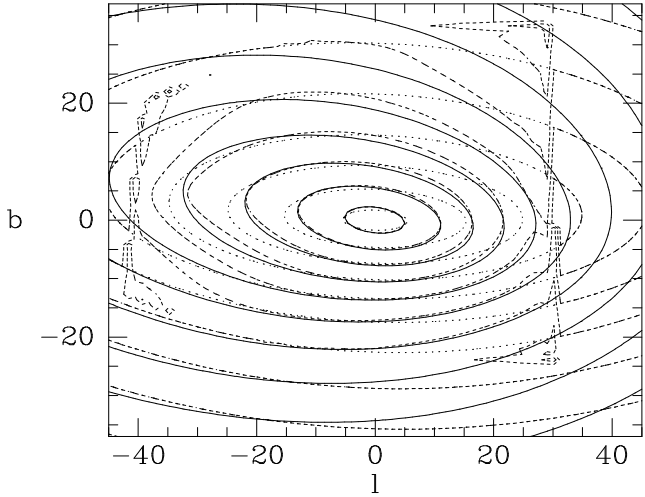

**Fig. 5** A trial in which $\theta_0 = \theta = 7°$, $\phi_0 = 30°$, $\phi = 18°$. The residuals tend to an rms value of $\ln(I/I_r) \simeq 0.4$ and a characteristic mismatch of projected and true isophotes is apparent. The true isophotes are shown by full curves and the model ones by dashed curves. The isophotes of the first (axisymmetric) model are shown dotted. This model had $(x_e, y_e, z_e) = (0.4, 0.4, 0.15)$ kpc.

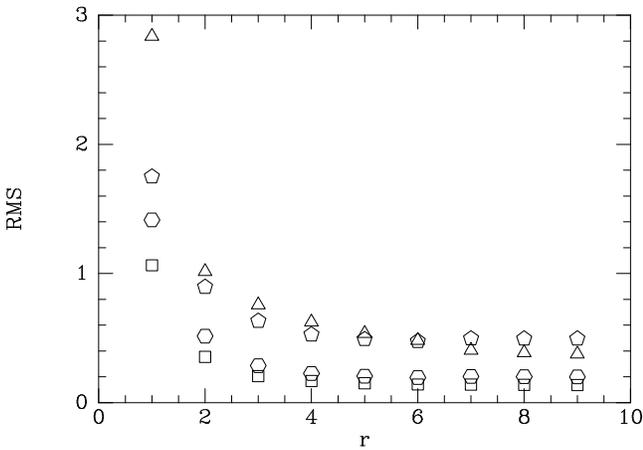

**Fig. 6** RMS residuals of $\ln(I/I_r)$ for the four deprojections shown in other figures. Triangles correspond to Fig. 1, squares to Figs 2 and 3, pentagons to Fig. 5 and hexagons to Fig. 7.

the $x$-axis, but at small $\theta$ the correct projected density can be produced with very non-elliptical isodensity surfaces, and for some reason the algorithm picks out such non-elliptical surfaces.

When the model is viewed from within a principal plane, the images of the symmetric companions that lie above and below the plane no longer provide independent information on the model's light distribution – the assumed symmetry of the model and data already suffice to ensure that if model and data agree at $b > 0$, then they must also agree at $b < 0$. Moreover, when $\theta = 0$, any motion of a source that leaves fixed on the sky the images of both the source and those of its symmetric companions that lie above the plane, will also leave fixed on the sky its symmetric companions below the plane. By contrast when $\theta \neq 0$, the symmetric companions below the plane will move on the sky even when those above the plane do not, and the amount by which they will move will increase continuously with $\theta$. Consequently, the information content of the data increases with $\theta$.

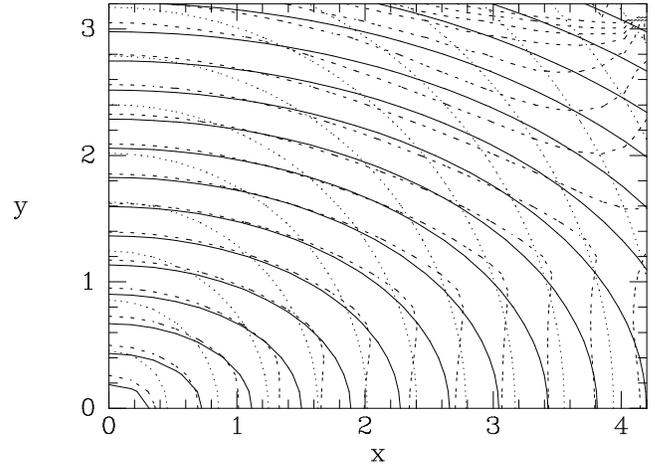

**Fig. 7** As for Fig. 3 except that the Sun-centre line lay in the equatorial plane ($\theta_0 = \theta = 0$, $\phi_0 = \phi = 18°$). The true and model isodensity surfaces intersect the equatorial plane in the full and dashed curves, respectively. The isodensity surfaces of the initial model generated the dotted curves. The dashed curves have unphysical shoulders that lie along a line that is inclined to the $x$-axis by an angle roughly equal to $\phi$.

As we have remarked, we cannot expect the deprojection to be unique even in the case $\theta \neq 0$, so it is not surprising that in the degenerate case $\theta = 0$ the algorithm picks a solution that differs materially from the source of the data. Fortunately, the spurious structure in the solution is readily identified because it is associated with the Sun–centre line: the difference between the original density and the recovered density consists largely of a cone of negative density around the $x$-axis and a ridge of positive density around the sun–centre line. This structure is reminiscent of axisymmetric konus densities (Gerhard & Binney 1995), which are intimately related to the non-uniqueness of the axisymmetric deprojection problem.

One can ensure the data are fitted by more nearly elliptical density distributions $\nu_r$ by restarting the algorithm from an elongated rather than a spherical distribution after one has discovered that the data require elongated contours in the $(x, y)$ plane. Alternatively, sharp features in the model can be discouraged by either filtering the model at each iteration or by biasing the Lucy iterations against them (Lucy 1994).

### 3.1 Noise

Since real data are always noisy, it is important to understand how robustly the algorithm copes with noise. Fig. 8 shows the results of deprojecting two noisy data sets. In the left panel the problem was identical to that of Figs 2 and 3 except that Gaussian noise was added to the data with dispersion $\sigma = 0.001I(0,0)$. This noise dominates the data for $l \gtrsim 20°$, but rapidly becomes unimportant further in. The left panel in Fig. 8 shows that the density inside 3 kpc is accurately recovered, and outside this radius the model rapidly becomes worthless. The right panel in Fig. 8 shows a similar experiment in which the true luminosity density had much larger scale lengths. Specifically, in this case the parameters $(x_e, y_e, z_e)$ of equation (11) were $(1.5, 0.9, 0.45)$ kpc. The noise level was again chosen such that noise and data were equal at $l \simeq 20°$. In this case noise is important even at $1-3$ kpc. Nonetheless, the right panel in Fig. 8 shows that after four iterations the algorithm has correctly diagnosed the elongation of the bar. In a practical case, this very noisy inversion would inspire one to repeat the analysis, this time starting from an appropriately elongated model distribution. After one or at most two iterations one's model would then provide a good fit to the data while still not being strongly contaminated by noise.

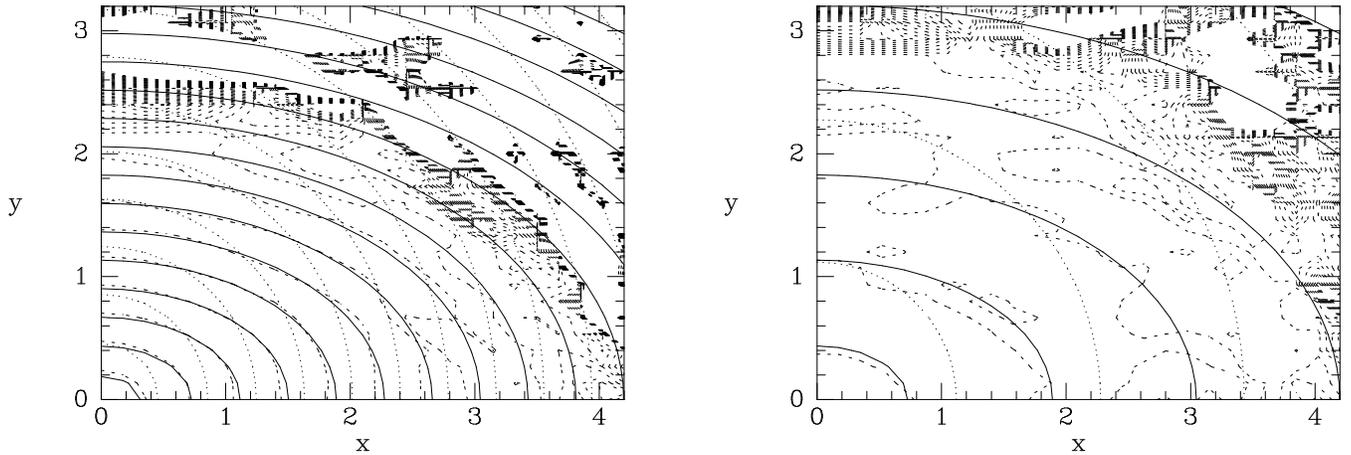

**Fig. 8** Deprojections of noisy data. The left panel is as for Fig. 3 except that Gaussian noise has been added to the data at a level that dominates the surface brightness beyond $l \simeq 20°$ – the dispersion was $0.001 I(0,0)$. The right panel shows the result of deprojecting a much shallower brightness distribution – the scale lengths in (11) were three times those used for the other figures. Again Gaussian noise $[\sigma = 0.07 I(0,0)]$ dominates the signal outside $l \simeq 20°$, but is now also important further in. These models were obtained after just four iterations with $\beta = 0.2$.

## 4  CONCLUSIONS

We have presented a Richardson–Lucy algorithm for the recovery of a mirror-symmetric density distribution from its image when viewed from a distance similar to our distance from the bulge of the Milky Way.

Practical application of this algorithm is complicated by the necessity of somehow artificially confining the three-dimensional model being recovered. We place the model in a box and use the simple rule (12), which depends on the initially guessed distribution $\nu_0$, to take into account emission from outside the box. Since the quality of $\nu_0$ soon manifests itself in the predicted distribution on the sky, it is in practice straightforward to iterate towards a good choice for $\nu_0$.

When the assumed orientation of the principal axes differs significantly from the true orientation, the algorithm correctly recovers the elongation of the isodensity surfaces, but the residuals on the sky remain relatively large. In such a case the residuals have a pattern on the sky indicative of the error made in the assumed principal-axis orientations.

Given a reasonable starting point $\nu_0$, the algorithm reliably detects non-axisymmetry in the "galactic plane." When the Sun-centre line does not lie too exactly in a principal plane of the model, the shape of the recovered isodensity surface is very good away from the edge of the bounding box. When the Sun-centre line lies nearly in a principal plane, the detailed shape of the recovered isodensity surfaces is liable to be unrealistically angular, although their elongation is correctly recovered.

The algorithm does not perform well when the data are very noisy. In general worthwhile results are obtained in regions for which the signal-to-noise is unity or greater. Fortunately, the quality of the results obtained at small radii, where the data may be expected to have high signal-to-noise, is not seriously undermined by the fact that every line of sight has to pass through low-density regions for which reliable data will not be available.

It will be interesting to see whether uncertainties due noise and dust absorption can be reduced to a level which allows this algorithm to be applied to the *COBE*/DIRBE near-infrared data.

## ACKNOWLEDGMENTS

This work was supported by NATO grant CRG 931469

Stark A.A., 1977, ApJ, 213, 368